# Large Amplitude Oscillatory Shear Study of a Colloidal Gel at the Critical State

Khushboo Suman,[1a] Sachin Shanbhag,[2*] and Yogesh M. Joshi[1*]

[1]Department of Chemical Engineering, Indian Institute of Technology Kanpur, Kanpur 208016, India

[2]Department of Scientific Computing, Florida State University, Tallahassee, Florida 32306 USA

[a]Present address: Department of Chemical & Biomolecular Engineering, University of Delaware, Newark, Delaware 19716 USA

* Author to whom correspondence should be addressed; electronic mail: sshanbhag@fsu.edu and joshi@iitk.ac.in

Abstract

We investigate the nonlinear viscoelastic behavior of a colloidal dispersion at the critical gel state using large amplitude oscillatory shear (LAOS) rheology. The colloidal gel at the critical point is subjected to oscillatory shear flow with increasing strain amplitude at different frequencies. We observe that the first harmonic of the elastic and viscous moduli exhibits a monotonic decrease as the material undergoes a linear to nonlinear transition. We analyze the stress waveform across this transition, and obtain the nonlinear moduli and viscosity as a function of frequency and strain amplitude. The analysis of the nonlinear moduli and viscosities suggests intracycle strain stiffening and intracycle shear thinning in the colloidal dispersion. Based on the insights obtained from the nonlinear analysis, we propose a potential scenario of the microstructural changes occurring in the nonlinear region. We also develop an integral model using the time-strain separable K-BKZ constitutive equation with a power-law relaxation modulus and damping function obtained from experiments. At low strain amplitudes, this model compares well with experimental data at all frequencies. However, a stronger damping function, which can be efficiently inferred using a spectral method, is required to obtain quantitative fits across the entire range of strain amplitudes and the explored frequencies.

## 1. Introduction:

The linear viscoelastic (LVE) properties of soft materials give crucial insights into the microstructure of the same.[1-5] Understanding how the latter affects the former is vital not just from an academic point of view but also from an industrial processing perspective.[6-8] Rheological behavior under nonlinear flow field associated with industrial processes cannot be described by merely the knowledge of linear viscoelastic response. The strains that a material is subjected to during a practical application can be quite large. Consequently, understanding microstructure specific linear – nonlinear viscoelastic transition in soft materials is of critical importance. In this work, we analyze linear – nonlinear viscoelastic transition of the space spanning percolated fractal network of colloidal particles (colloidal gel at the critical state) using large amplitude oscillatory shear (LAOS) approach.

The investigation of soft materials by applying large amplitude oscillatory shear dates back to early 1960s.[9] The measurement of a response (the stress or the strain) was conducted in an analog fashion, where the nonlinearity was represented by a deviation in the closed-loop plots (Lissajous-Bowditch curves) of stress versus strain or stress versus strain rate.[10, 11] The graphical way of representation of the measured data by Lissajous-Bowditch curves has been a major characterization feature of LAOS.[10, 12] Matsumoto *et al.*[13] were the first to quantify the nonlinearity in terms of the third harmonic of moduli for a particulate suspension in polystyrene solution. Giacomin and Oakley[14] further analyzed the Lissajous-Bowditch curve to obtain the Fourier series from it. The Fourier transform analysis of the stress data had become one of the prevailing methods of quantifying the nonlinear viscoelasticity.[11, 13, 15-17] This led to the birth of the term 'Fourier transform (FT) rheology.'[18] Since then, it has been routinely used to study the nonlinear rheological behavior of soft materials.[19-22] While the FT rheology framework is mathematically robust, it fails to provide a physical interpretation of the higher harmonic coefficients.[23] In order to overcome this shortcoming, Cho *et al.*[24] developed the stress decomposition (SD) method, wherein the total stress is decomposed into elastic and viscous components. The SD methodology has been proven to be mathematically equivalent to FT rheology.[25] Ewoldt *et al.*[26] replaced the unknown material functions in the SD method by orthogonal Chebyshev polynomials and coefficients. Later, Hyun and Wilhelm[27] introduced a nonlinear parameter known as the $Q$ parameter, which is expressed in terms of the ratio of intensities of first and third harmonic. The shape of the $Q$ curve has been analyzed to distinguish between linear and branched polymers.[5]

Rogers and coworkers[23, 28] suggested a new qualitative analysis and described the response of yield stress materials as a sequence of physical processes (SPP) by introducing the local defined measures including time dependent dynamic moduli. However, detailed information at every time step throughout the cycle is required for SPP to describe the material response.[23, 29] Furthermore, SPP approach is a high-dimensional representation even in the linear viscoelastic regime and is observed to inaccurately report elasticity for a generalized Newtonian fluid model, which makes it a challenging approach.[29, 30] Recently Erturk *et al.*[31] compared SPP approach and Chebyshev polynomials approach to analyze nonlinear viscoelastic behavior of wheat flour dough and pectin solution. Over past two decades various approaches have been employed to study LAOS response to various kinds of materials. How LAOS parameters can be analyzed to obtain meaningful information about microstructure, and how it gets affected by nonlinear deformation has been reviewed by Hyun *et al.*[32] and more recently by Cho.[18] Lately there has been a greater thrust on analyzing nonlinear complex moduli, which depend only frequency.[33-36] The nonlinear complex moduli can be obtained by expressing the stress waveform in response to sinusoidal strain input as a power series.[37, 38] For medium amplitude oscillatory shear, Ewoldt and coworkers[23] analyzed nonlinear complex moduli and proposed that their signs give physically meaningful interpretations. However, experimental measurement of these intrinsic parameters requires fine resolution and high precision at low torque limits.

In this work, we study LAOS behavior of the unique state called critical state in a colloidal gel. Sol-gel transition, wherein material transforms from a free-flowing liquid state to a non-flowing state with network-like microstructure, has been observed for polymeric[39, 40] as well as colloidal systems.[41, 42] Such sol-gel transition could be spontaneous,[43] because of change in temperature[39] or concentration of one of the species.[44] It has been observed that in a limit of free flowing liquid state, elastic ($G'$) and viscous ($G''$) moduli show a terminal behavior given by: $G' \sim \omega^2$ and $G'' \sim \omega$, where $\omega$ is the probed frequency associated with the linear response regime.[45] As the extent of crosslinking increases, $\omega$ dependence of both the moduli decreases, with that asscciated with $G''$ decreasing at a faster rate. Eventually growing clusters connect with each other in such a fashion that they just form a space spanning percolated network. Such weakest network has been represented as a critical gel state in the literature.[45] The critical gel has a unique signature wherein $G'$ and $G''$ are observed to show an identical power law dependence on $\omega$ given by:[45]

$$G' = G'' \cot(n_c \pi/2) = \frac{\pi S}{2\Gamma(n_c)\sin(n_c \pi/2)} \omega^{n_c}, \tag{1}$$

where $n_c$ is the power law index ($0 < n_c < 1$), $S$ is the gel strength, while $\Gamma(...)$ represents the Euler Gamma function. Knowledge of the power law index $n_c$ leads to fractal dimension ($f_d$) of the percolated network $f_d = 5(2n_c - 3)/2(n_c - 3)$.[46] Remarkably this relationship between $n_c$ and $f_d$ has been validated for fractal gels that were subjected to simultaneous rheology and scattering studies.[47, 48] Nonlinear viscoelastic behavior of the critical gels has been investigated by performing step strain experiments at different magnitudes of strains in the linear and nonlinear regime.[49, 50] The resultant self-similar nature of the relaxation modulus leads to damping function, which has been used to analyze linear – nonlinear transition in the critical gels. Suman and Joshi[50] reported strain softening for colloidal clay gel while Keshavarz et al.[49] reported the strain hardening followed by strain softening behavior for casein network formed through physical interactions. Nonlinear rheological behavior at the critical state has also been studied for a food gels[51] as well as a polymer gel.[52] Gisler et al.[53] investigated diffusion limited aggregate of fractal colloidal gel and reported strain hardening originating from the rigid backbone of the same. However, a comprehensive experimental characterization and theoretical modeling of the nonlinear viscoelastic behavior of the uniquely defined critical gel state is still in its infancy.

In this work, we present a detailed analysis of nonlinear measurements of a colloidal dispersion of synthetic hectorite clay at the critical gel state. Based on our observations, we propose a picture of the microstructural changes occurring during the transition from linear to nonlinear regime. We also model the nonlinear behavior using an integral constitutive equation that uses information of linear viscoelasticity and damping function obtained from nonlinear stress relaxation modulus. The proposed model with further modification of the damping function predicts the intracycle stress response in the linear and nonlinear region quite well, and provides physical insights during the nonlinear transition.

## 2. Large Amplitude Oscillatory Shear: Theory

In order to characterize the linear and nonlinear response, the material is subjected to a sinusoidal strain input with amplitude $\gamma_0$ at a constant frequency ($\omega$) given by:

$$\gamma = \gamma_0 \sin \omega t. \tag{2}$$

The linear stress response on the application of a sinusoidal strain input can be represented as:[54]

$$\sigma = \gamma_0 [G' \sin(\omega t) + G'' \cos(\omega t)], \tag{3}$$

where $G'$ and $G''$ are elastic and viscous moduli respectively. Furthermore, on eliminating time variable in the above equations, we get the direct relationship between stress and strain given by:

$$\sigma^2 - 2G'\sigma\gamma + \gamma^2 (G'^2 + G''^2) = (G''\gamma_0)^2. \tag{4}$$

The above equation suggests the Lissajous-Bowditch curve (stress-strain curve) corresponds to a straight line for a linear elastic material where stress is in phase with strain. On the other hand, stress is out of phase with strain for purely viscous material and the Lissajous-Bowditch curve is represented by a circle. Importantly, for a linear viscoelastic material, the Lissajous-Bowditch curve is elliptical.

In the nonlinear region, $G'$ and $G''$ are not uniquely defined since stress response is not a single harmonic sinusoid.[55] The most common method to access the nonlinearity is Fourier transform rheology.[56] The stress response on the application of a sinusoidal strain input can be represented by a Fourier series as:[54]

$$\sigma(t;\omega,\gamma_0) = \gamma_0 \sum_{n,odd} [G'_n(\omega,\gamma_0) \sin(n\omega t) + G''_n(\omega,\gamma_0) \cos(n\omega t)], \tag{5}$$

where $G'_n$ and $G''_n$ are the $n^{th}$ order elastic and viscous moduli respectively. The Fourier series contains only odd harmonics in order to obey the directionality of strain, which suggests that on reversing the coordinate system, the material response remains unchanged. The stress response consists of first harmonic only in the limit of linear viscoelasticity and the material properties are characterized by $G'$ and $G''$, while the higher harmonic appears with an increase in the strain amplitude in the nonlinear region. Even though the FT framework is mathematically complete, it fails to provide a physical interpretation of the higher-order coefficients.

In order to interpret the LAOS data meaningfully, Ewoldt et al.[26] adopted the method of orthogonal stress decomposition proposed by Cho et al.[24] The idea was to decompose the stress response into a summation of elastic stress $(\sigma')$ and viscous stress $(\sigma'')$ using symmetry arguments as:

$$\sigma'(\gamma) = \gamma_0 \sum_{n,odd} G'_n(\omega, \gamma_0) \sin(n\omega t), \text{ and} \qquad (6)$$

$$\sigma''(\gamma) = \gamma_0 \sum_{n,odd} G''_n(\omega, \gamma_0) \cos(n\omega t). \qquad (7)$$

The analysis of the nonlinear parameters underscores the microstructural changes occurring in the material on application of the deformation field in the nonlinear domain.

3. **Material and Experimental Procedure:**

In this work, we prepare a dispersion of 3 weight % synthetic hectorite clay LAPONITE XLG® having 3 mM NaCl. Oven dried (120°C for 4 h) white powder of hectorite was added to ultrapure millipore water with added salt, and stirred using ultra Turrax drive for 30 minutes. The detailed protocol to prepare the dispersion is discussed in our previous publication.[50] Hereafter, we refer to this dispersion as hectorite dispersion. The freshly prepared dispersion was loaded in the sandblasted concentric cylinder geometry having a cup diameter of 30.38 mm and a gap of 1.17 mm on TA Instruments DHR 3 rheometer for rheological measurements. Subsequent to the loading of the sample, it was subjected to cyclic frequency sweep within the linear viscoelastic regime with a stress magnitude of 0.1 Pa and frequency range of 0.5-25 rad/s. It takes 100 s to complete one frequency sweep. The amplitude sweep and all the nonlinear measurements were performed at the critical gel state of the colloidal dispersion. In large amplitude oscillatory shear (LAOS) test, the critical gel is subjected to a sinusoidal strain of the form: $\gamma = \gamma_0 \sin \omega t$ where $\omega$ is a constant frequency and $\gamma_0$ is the amplitude of strain varying logarithmically from 0.001 to 10. The LAOS tests were performed at various frequencies of 0.1 Hz, 0.5 Hz and 5 Hz. We did not explore higher strain amplitudes and frequencies to avoid fracture of sample at the free surface. The transient data was collected using the rheometer software Trios. For every strain input, the raw values of 900 data points were recorded in each cycle. The smoothing of the data and analysis of the stress and strain waveform data was performed using MITlaos program (Version 2.2 beta).[57] The elastic and viscous stress components, Chebyshev coefficients, higher harmonic viscoelastic moduli and Fourier spectrum were determined using MITlaos. The Fourier coefficient obtained from MITlaos program were also verified independently by performing Fourier transform using Matlab. All the experiments were conducted at a constant temperature of 30°C maintained by a Peltier plate temperature system. We employ a solvent trap and also add a thin layer of silicone oil to the free surface of the sample to prevent the evaporation of water during the measurements.

4. **Results and Discussions:**

The freshly prepared colloidal dispersion is subjected to cyclic frequency sweep, which enables the estimation of viscoelastic properties at different extents of gelation. In Figure 1 (a), we plot the evolution of elastic $(G')$ and viscous $(G'')$ modulus with respect to angular frequency $(\omega)$ during the process of sol-gel transition. The modulus values at the respective times have been shifted vertically for clarity and the corresponding shift factors are mentioned in the legend. At very small times, $G''$ dominates $G'$ and the dispersion exhibits characteristic terminal behavior with $G' \sim \omega^2$ and $G'' \sim \omega$ dependence, which is suggestive of the sol state. In Figure 1 (b) we plot the evolution of loss tangent $(\tan\delta)$ as a function of time at different frequencies, which also shows a prominent decrease with $\omega$. However, as time progresses, both the moduli grow with the rate of growth of $G'$ being higher than $G''$. Eventually, $G'$ surpasses $G''$ and the dependence of $G'$, $G''$ and $\tan\delta$ on $\omega$ weakens with an increase in time. At a particular time, both the moduli exhibit identical dependence on $\omega$ $(G' \sim G'' \sim \omega^{0.27})$ while $\tan\delta$ becomes completely independent of the applied frequency. Such dependence is the characteristic rheological signature of the critical gel transition. The time to achieve the critical gel transition is indicated by a solid arrow in Figure 1 (b). Very interestingly, the critical relaxation exponent $(n_c)$ computed at the critical gel state using the relation $n_c = 2\delta/\pi$ is identical to the power-law exponent of viscoelastic moduli on $\omega$ (As suggested by Eq. 1). At further higher times, $G'$ and $G''$ are only weakly frequency-dependent while the dependence of $\tan\delta$ on $\omega$ inverts and eventually $\tan\delta$ increases with an increase in $\omega$. This is a distinctive feature associated with the post-gel state. Therefore, Figure 1 clearly suggests that the results exhibited by a freshly prepared dispersion of synthetic hectorite clay are consistent with a system undergoing sol-gel transition.

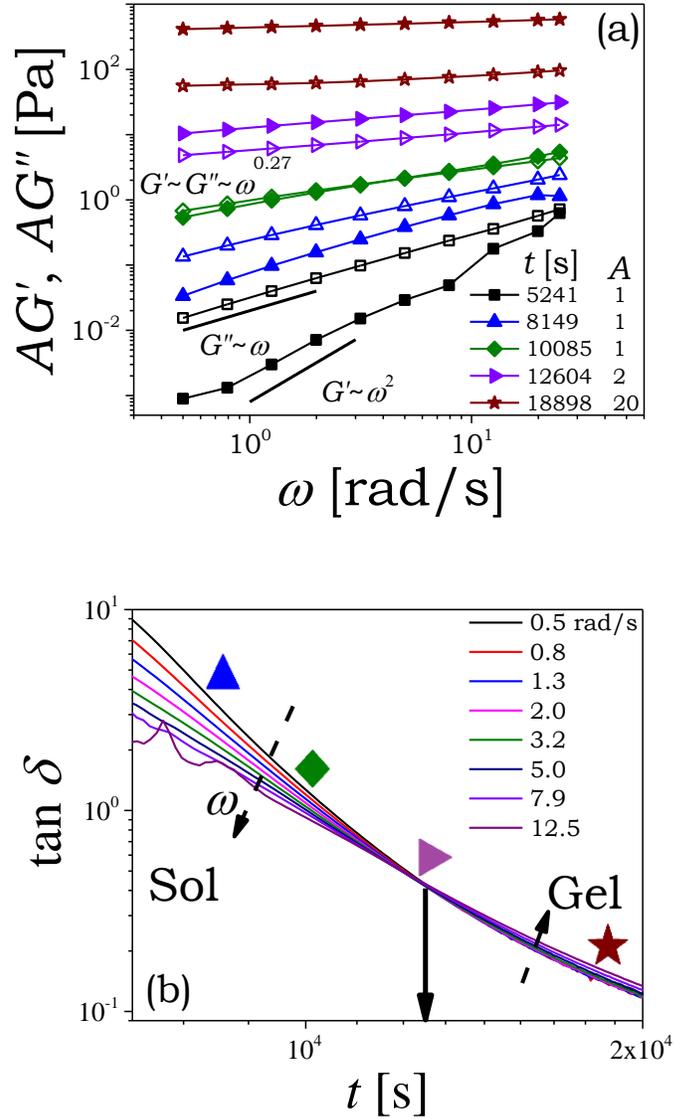

**Figure 1.** (a) The variation in elastic modulus ($G'$, closed symbols) and viscous modulus ($G''$, open symbols) is plotted as a function of angular frequency ($\omega$) at the indicated times in the legend. The value of modulus on the vertical axis has been shifted by a factor $A$ as mentioned in the legend for the purpose of clarity. The lines serve as a guide to the eye. (b) The evolution of $\tan\delta$ is plotted as a function of time during the process of sol-gel transition in 3 weight % synthetic hectorite clay dispersion with 3mM NaCl. The dashed arrow indicates the direction of the increase in the value of frequency. The solid arrow denotes the point of critical gel transition. The time instances indicated in figure (a) are shown by the same symbol in figure (b).

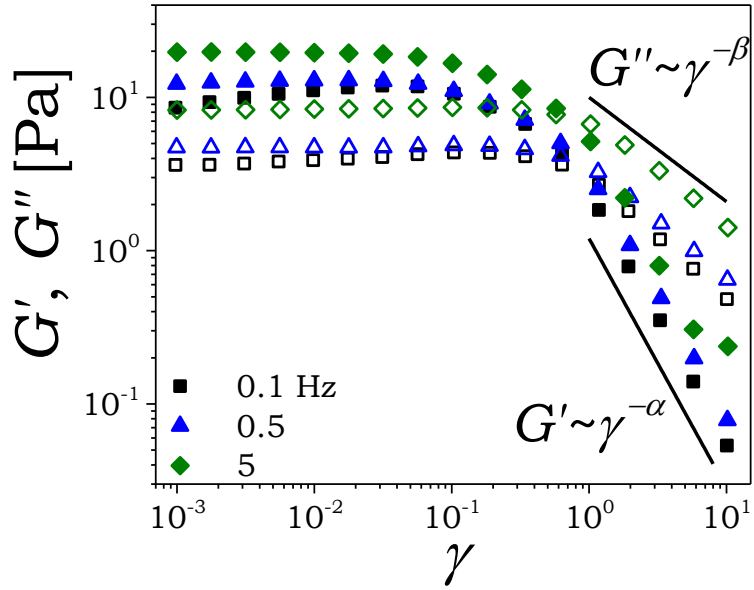

**Figure 2.** The variation of $G'$ (closed symbols) and $G''$ (open symbols) at the critical gel state is plotted as a function of strain amplitude at different magnitudes of probed frequency. The solid lines denote the power-law dependence of the moduli.

We now aim to study the nonlinear rheological behavior of the material at the critical gel state. In order to ensure the colloidal dispersion is at the critical gel state, we first subject the freshly prepared dispersion to cyclic frequency sweep experiment. At the point when $\tan\delta$ becomes frequency independent, we subject the critical gel to an oscillatory strain sweep. The critical gel state is the weakest space spanning fractal network as it is on the verge of a pre-gel state. While densification of the network in hectorite dispersion is a continuous process, the network can be considered to closely resemble the critical gel state for a duration of at least 100s before it enters the post gel state.[50] In Figure 2, we plot the variation in $G'$ and $G''$ as a function of strain amplitude $(\gamma)$ at three different values of explored frequencies. The frequency values are chosen in such a way that the system remains in the critical gel state over the complete duration of strain sweep. At low strain amplitude, as expected, the critical gel exhibits constant moduli. The value of $G'$ is greater than $G''$ in the linear viscoelastic region, by virtue of $n < 0.5$, thus indicating the critical gel to be dominated by elastic effects. In this regime, $G'$ and $G''$ increases with an increase in the magnitude of the applied frequency. The linear viscoelastic regime is observed till $\gamma \approx 0.1$ for all the frequencies after which the value of $G'$ starts to decrease as the critical gel enters the nonlinear viscoelastic region. It can be seen in Figure 2 that both $G'$ and $G''$ decrease with an increase in the strain amplitude in the nonlinear regime. Since the critical gel state is the weakest percolated network, the energy dissipated during the linear – nonlinear transition is less, which does not lead to any peak in $G''$.[58] In the nonlinear regime, $G'$ and $G''$ lose their original meaning, and represent $G'_1$ and $G''_1$. In a limit of large $\gamma$, the decay in $G'_1$ and $G''_1$ follows a power law given by $G'_1 \sim \gamma^{-\alpha}$ and $G''_1 \sim \gamma^{-\beta}$. The power-law exponents are observed to be independent of the applied frequency and takes the value as $\alpha = 1.62 \pm 0.015$ and $\beta = 0.74 \pm 0.035$ for all the explored values of frequency. Similar value of power-law exponents has been reported for Hectorite-PEG dispersion[59] and pluronic-hyaluronic acid gels.[22] The critical gel shows a similar qualitative response at the explored frequencies.

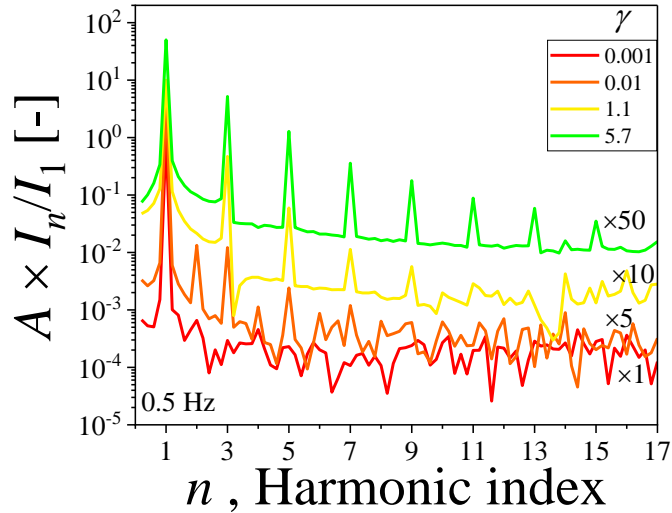

**Figure 3.** The normalized Fourier transform spectra at a primary frequency of $f$ = 0.5 Hz is plotted as a function of harmonic index *n* for different strain amplitudes mentioned in the legend. The individual spectrum has been vertically shifted by an amount *A* mentioned on the plot.

The transient data collected during an oscillatory strain sweep is analyzed to obtain various linear and nonlinear viscoelastic parameters. We first analyze the transient data using Fourier Transform (FT) Rheology. The viscoelastic nonlinearity can be identified by the appearance of high-order harmonics on the Fourier transform spectra of the stress in the frequency domain. In Figure 3, we plot the intensity of the *n*th harmonic related to the primary one ($I_n / I_1$) at various strain amplitudes ranging from the linear to nonlinear region with a primary frequency of 0.5 Hz. The noise floor appears near a normalized intensity of $I_n / I_1 \approx 2 \times 10^{-4}$. For smaller strain amplitudes belonging to the linear viscoelastic region ($\gamma = 0.001$ and $0.01$), there is a primary peak at $n = 1$ while all higher harmonics are negligible ($I_n / I_1 \approx 2 \times 10^{-3}$). For the strain of $\gamma = 10^{-2}$ the peak associated with the second harmonic can be seen to be as strong as that of third harmonic. However, both the peaks are around two orders of magnitude smaller than that associated with the primary harmonic indicating the nonlinear effects to be too weak. Such low value of even harmonics confirms the absence of wall slip[60, 61] or secondary flows.[62] With increase in the strain amplitude to $\gamma = 1.1$, the intensity peaks at higher odd harmonics become dominant, thus signifying the transition into the nonlinear region. At the highest strain amplitude of $\gamma = 5.7$, the nonlinearity is more pronounced by the peaks at high odd harmonics.

In Figure 4 (a) we plot the variation of $G'$ and $G''$ as a function of strain amplitude at $f = 0.5$ Hz. The linear to nonlinear transition can be seen in Figure 4 (a) with an increase in the strain magnitude. In Figure 4 (b), we show the intracycle response by plotting the total stress as a function of strain for the strain magnitudes highlighted in Figure 4 (a) with identical symbols. It can be seen in Figure 4 (b), that the stress induced in the critical gel at low amplitudes of strain is small, which as expected increases with an increase in the strain amplitude. The response of the critical gel to the strain loading in the linear viscoelastic region is primarily elliptical as suggested by Eq. (4). Furthermore, the area enclosed by the stress-strain closed curves increases with increasing strain amplitude. At large strain amplitude as the system gets into the nonlinear regime, the shape of the stress-strain curve becomes qualitatively different. Using the framework discussed by Cho *et al.*[24] and Ewoldt *et al.*,[26] the total stress can be decomposed into the elastic and viscous contributions which are shown in Figure 4 (c) and (d) respectively. While the total stress response is a hysteresis loop in the stress-strain plane, the elastic and viscous stresses are single-valued functions, when plotted respectively as a function of $\gamma$ and $\dot{\gamma}$. It can be seen in Figure 4 (c) and (d) that with an increase in the strain amplitude, the nonlinearity is evident through the deformation of the stress-strain curves. While the elastic stress curves plotted against $\gamma$ turn upwards, the viscous stress plotted as a function of $\dot{\gamma}$ can be seen to curve downwards at the highest amplitude of strain.

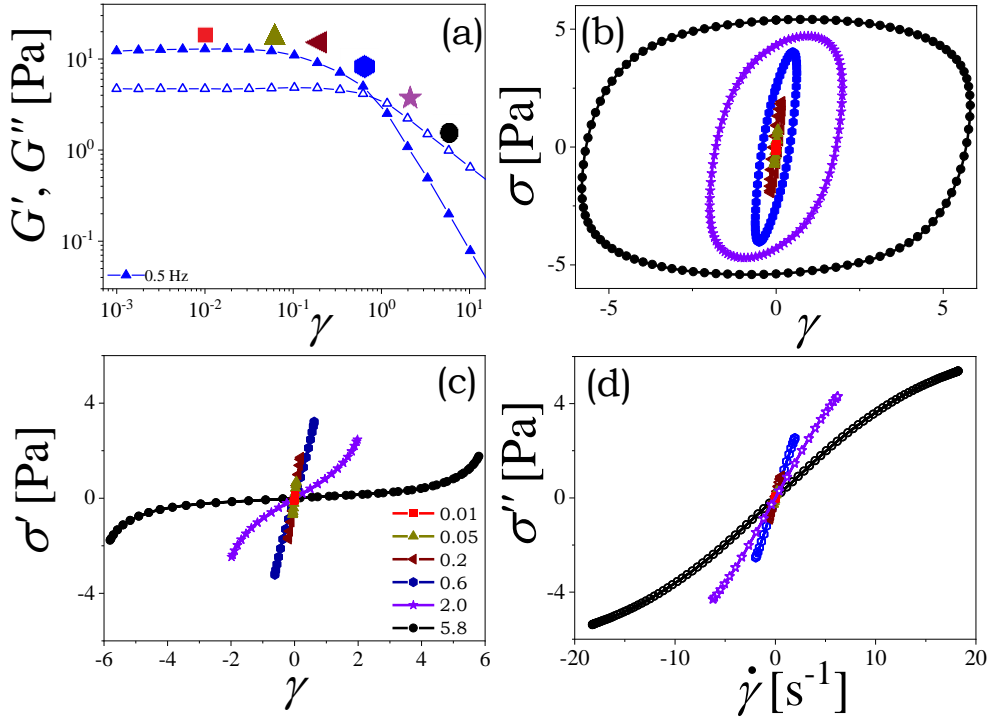

**Figure 4.** (a) The variation of $G'$ (closed symbols) and $G''$ (open symbols) is plotted as a function of strain amplitude at $f = 0.5$ Hz. (b) The Lissajous-Bowditch curves are plotted at different amplitudes of strain for the same frequency. (c) Elastic stress ($\sigma'$) is plotted as a function of $\gamma$ and (d) Viscous stress ($\sigma''$) is plotted as a function of $\dot{\gamma}$ for different strain amplitudes. In figures (b), (c) and (d) the corresponding values of strain are indicated by the identical symbols in Figure (a).

The nonlinearities can also be examined graphically from the 3D Lissajous-Bowditch plots in Figure 5. The projections of the Lissajous-Bowditch curves on stress-strain, stress-shear rate and strain-shear rate planes are shown in Figure 5. In an oscillatory cycle, the variation of strain is orthogonal to the strain rate. Consequently, the strain achieves the maximum value when the strain rate is zero and the other way around. While the stress-strain curves are elliptical in the linear viscoelastic region, the shape deviates from a typical elliptical response as the flow field enters the nonlinear regime. As the stress-strain Lissajous-Bowditch curve approaches a rectangular shape at high amplitude of strain field, the stress-strain rate Lissajous-Bowditch curve bend over at maximum strain rate. It is interesting to note that even though the oscillatory strain sweep does not show any variation in the viscoelastic properties in the linear domain, the Lissajous-Bowditch curve associated with each strain amplitude in the linear domain is different. The unique shape associated with all the strain amplitudes sets an unambiguous measure of quantifying the linear as well as the nonlinear response of the colloidal dispersion. Therefore, the Lissajous-Bowditch curves act as a rheological fingerprint test of a material.

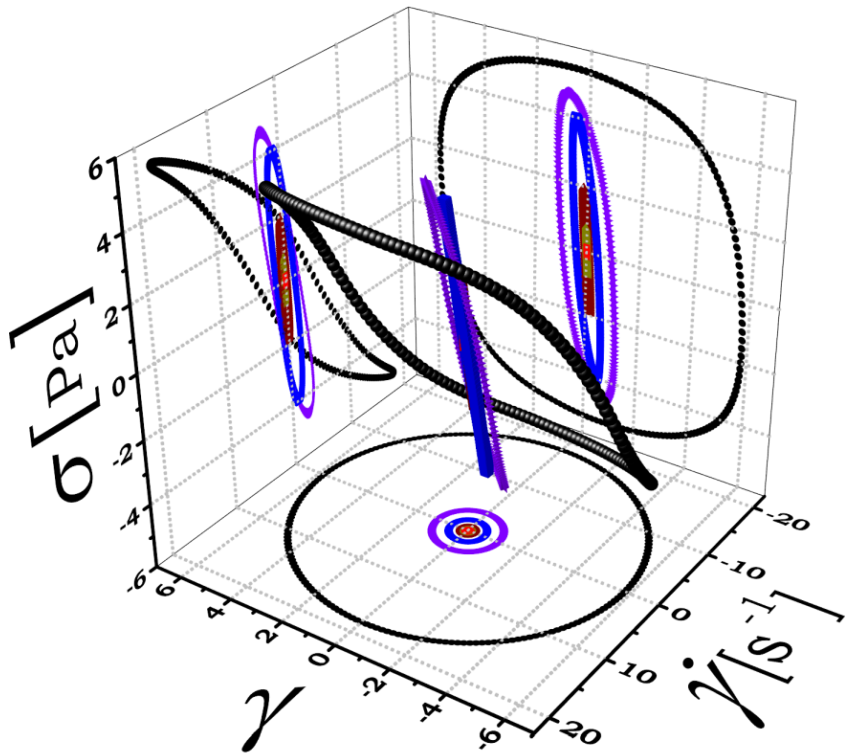

**Figure 5.** The 3D Lissajous-Bowditch curves and its projections on the stress-strain, stress-shear rate and strain-shear rate planes are plotted at $f = 0.5$ Hz. The legend in this figure is same as that mentioned in Figure 4 (c).

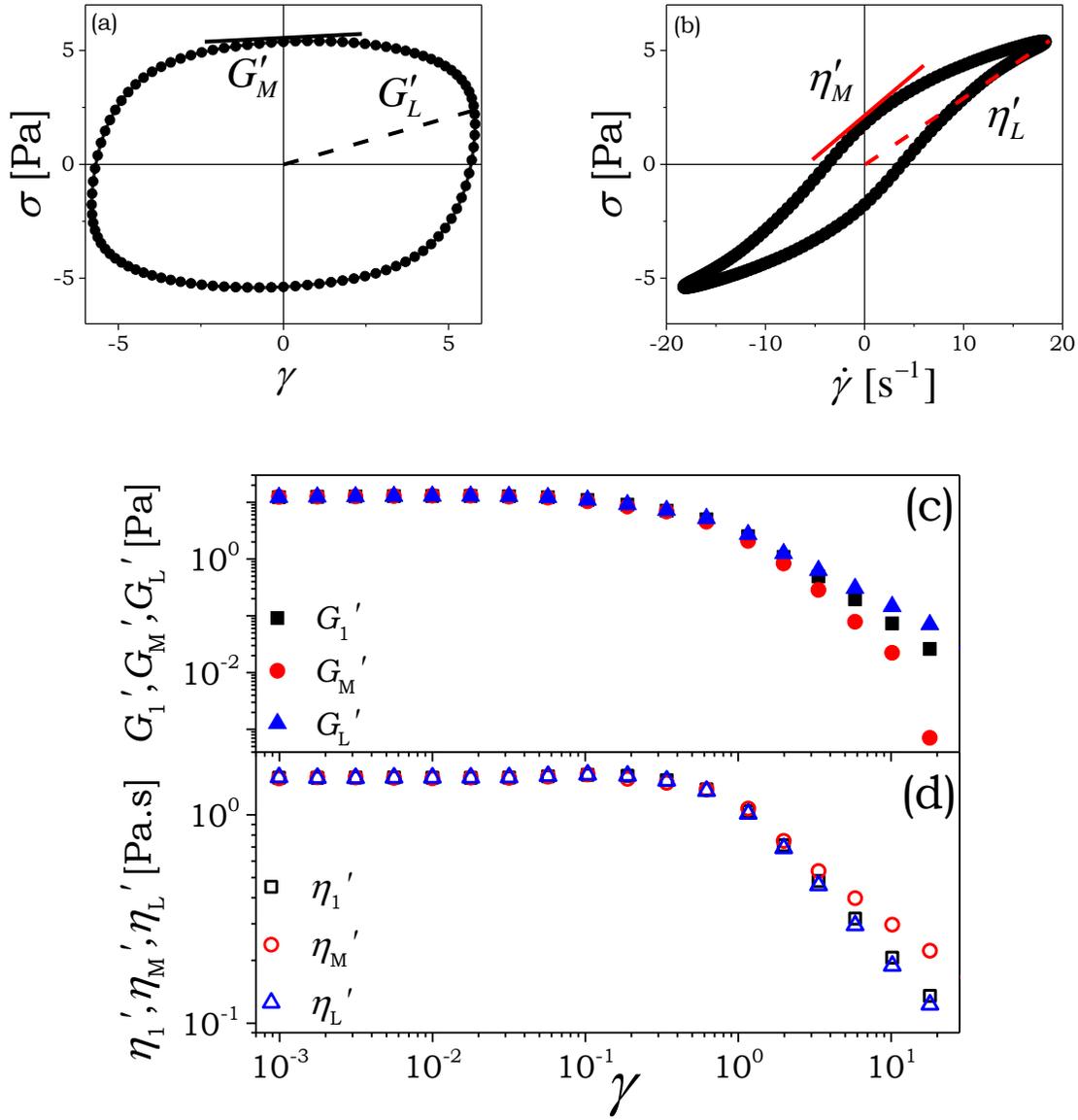

**Figure 6.** Representation of the (a) nonlinear elastic moduli $G'_M$ (solid line) and $G'_L$ (dashed line) (b) nonlinear viscosities $\eta'_M$ (solid line) and $\eta'_L$ (dashed line) in terms of the slope of Lissajous-Bowditch curves at $\gamma = 5.7$ and $f = 0.5$ Hz. The variation of (c) $G'_1$, $G'_M$, and $G'_L$ (d) $\eta'_1$, $\eta'_M$, and $\eta'_L$ is plotted as a function of strain amplitude at $f = 0.5$ Hz.

In order to analyze the shape of the Lissajous-Bowditch curve and account for the nonlinearities in the stress response, the following new parameters are defined. The minimum-strain ($G'_M$) and large-strain moduli ($G'_L$) are defined as:[26]

$$G'_M \equiv \left.\frac{d\sigma}{d\gamma}\right|_{\gamma=0} = \sum_{n,odd} nG'_n, \tag{8}$$

$$G'_L \equiv \left.\frac{\sigma}{\gamma}\right|_{\gamma=\pm\gamma_0} = \sum_{n,odd} G'_n(-1)^{\frac{(n-1)}{2}}. \tag{9}$$

These moduli are defined in such a way that each of the moduli reduces to $G'$ in the limit of linear viscoelasticity, which, in principle, represents the first harmonic of elastic modulus in the linear viscoelastic domain ($G'_1$). The graphical interpretation of nonlinear moduli at a strain amplitude in the nonlinear region is shown in Figure 6 (a). It can be seen that $G'_M$ is the tangent modulus at minimum strain amplitude while $G'_L$ is the secant modulus at maximum strain amplitude. The nonlinear moduli quantify the extent of distortion within a cycle. The first harmonic of elastic modulus ($G'_1$), minimum strain modulus ($G'_M$) and large strain modulus ($G'_L$) are plotted in Figure 6 (c) as a function of strain amplitude. At low amplitudes of strain (linear viscoelastic region), both the moduli $G'_M$ and $G'_L$, reduce to $G'_1$, where a Lissajous-Bowditch curve has a perfectly ellipsoidal shape. As expected, the increase in the magnitude of strain, at the onset of nonlinearity, results in a decrease in $G'_M$ and $G'_L$, such that in the nonlinear region, $G'_M$ and $G'_L$ are found to deviate dramatically from $G'_1$. This deviation of $G'_M$ and $G'_L$ from $G'_1$ signifies the divergence of a Lissajous-Bowditch curve from a linear viscoelastic response. The increasing extent of deviation of nonlinear moduli from $G'_1$ with increase in strain amplitude suggests a greater extent of distortion in the Lissajous-Bowditch curves. Furthermore, $G'_M$ decreases at a faster rate than $G'_L$ as evident at higher values of strain amplitude. This suggests that within a particular strain cycle at constant strain amplitude, $G'_L$ is greater than $G'_M$. The dominance of $G'_L$ over $G'_M$ suggests that the system is able to store more energy at the highest amplitude of strain than at the minimum value of strain. This advocates towards intracycle strain stiffening behavior in the critical gel state.[30] However, it is important to note that, overall, the value of moduli is decreasing with an increase in the strain amplitude, which suggests overall strain thinning behavior in the critical gel state.

In order to quantify the dissipative response of the material in the nonlinear domain, a set of dynamic viscosities have also been defined as:[26]

$$\eta'_M \equiv \frac{d\sigma}{d\dot{\gamma}}\bigg|_{\dot{\gamma}=0} = \frac{1}{\omega}\sum_{n,odd} nG''_n(-1)^{(n-1)/2}, \quad (10)$$

$$\eta'_L \equiv \frac{\sigma}{\dot{\gamma}}\bigg|_{\dot{\gamma}=\pm\dot{\gamma}_0} = \frac{1}{\omega}\sum_{n,odd} G''_n, \quad (11)$$

where $\eta'_M$ is the minimum-rate dynamic viscosity and $\eta'_L$ is the large-rate dynamic viscosity. In the limit of linear viscoelasticity, both the viscosities reduce to $\eta'_M = \eta'_L = \eta'_1 = G''_1/\omega$. The nonlinear viscosities are graphically represented in Figure 6 (b). The viscous nonlinearities are plotted as a function of strain amplitude in Figure 6 (d). As expected, $\eta'_M$ and $\eta'_L$ remain identical to $\eta'_1$ in the linear viscoelastic region. In the nonlinear region, $\eta'_M$ and $\eta'_L$ decreases with the rate of decay of $\eta'_L$ being higher than $\eta'_M$. This suggests that the value of $\eta'_M$ is always higher than $\eta'_L$ in the nonlinear region. Such a dependence of higher viscosity at minimum strain rate than at highest strain rate is an indicator of intracycle shear thinning behavior of the critical gel state.[26] We also analyze behavior of Chebyshev coefficients as a function of amplitude of strain amplitude at various exploted frequencies. The corresponding behavior is discussed in the Supplementary Material. The inference of the Chebyshev coefficients analysis matches well with that reported by Figure 6.

The elastic Lissajous-Bowditch curves at eight different strain amplitudes and varying frequency represented by solid black lines are shown as a Pipkin diagram in Figure 7. This diagram is defined using two independent parameters: the strain amplitude on the vertical axis and the frequency on the horizontal axis. Each Lissajous-Bowditch curve within the Pipkin diagram represents a plot of normalized stress on the vertical axis with normalized strain on the horizontal axis. The normalization of strain is carried out using the amplitude of imposed strain ($\gamma_0$ indicated on the vertical axis of Pipkin space) and stress is normalized using the maximum value of stress achieved in each cycle (indicated at bottom right corner of or inside each Lissajous-Bowditch curve). The local shape of the Lissajous-Bowditch curves depends on all the harmonics present in the system. For the linear viscoelastic region in Figure 7 ($\gamma < 0.19$), the stress-strain curves are elliptical at all the explored frequencies. As strain amplitude increases ($\gamma > 0.19$), Lissajous-Bowditch curves deviate from an elliptical shape, and get distorted as higher-order harmonics become non-zero. The elastic Lissajous-Bowditch curves broaden in the center, and the area enclosed by a Lissajous-Bowditch curve increases with an increase in strain amplitude. The qualitative effects of increasing strain amplitude are similar at all the explored frequencies. Furthermore, the maximum stress attained in each cycle increases as the oscillation frequency increases. The weak change in Lissajous-Bowditch curves upon increase in the frequency is a consequence of the weak increase of the moduli with frequency with a power-law exponent of 0.27 for the critical gel state. Such weak change in Lissajous-Bowditch curves across the frequencies is also reported in critical gel of gluten,[63] and colloidal glass of star-like micelles system.[64] At the highest explored frequency and strain amplitude of 5.7 at 5 Hz, the box-shaped Lissajous-Bowditch curve displays a preferred left inclination. Such feature is also observed in colloidal glass,[64, 65] gels[42] and polymer solutions[66] but unfortunately it is not very well understood in the literature. Further investigation at high strain and frequency in understanding the physical mechanisms responsible for such curious behavior is an important topic for future study.

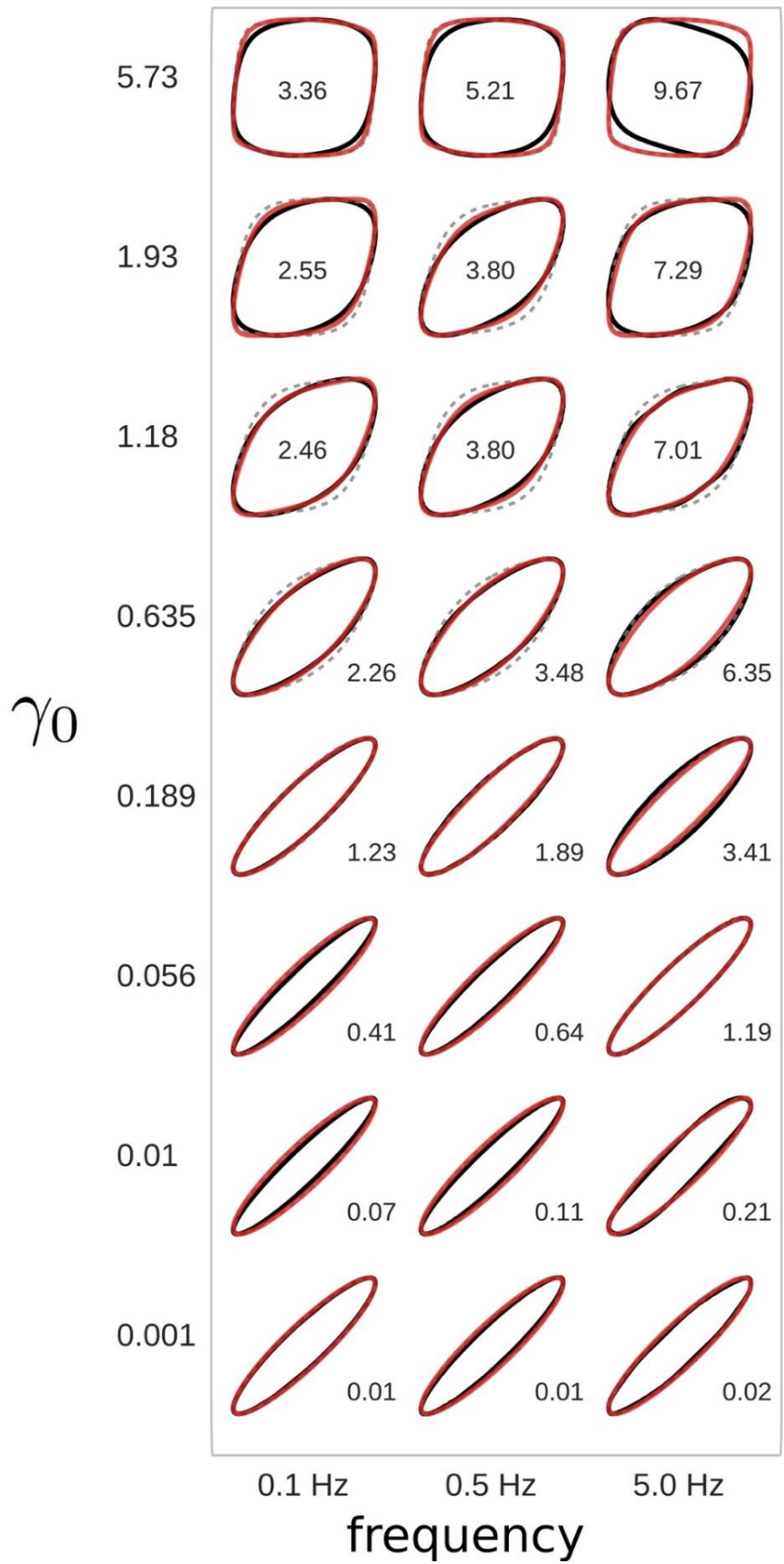

**Figure 7.** The normalized elastic Lissajous-Bowditch curves (solid black lines) are presented in a Pipkin space spanning frequency in the range of 0.1 Hz to 5 Hz and strain amplitude from 0.001 to 5.7. The dashed gray lines are predictions of TSS-KBKZ constitutive model using damping function parameters obtained from step-strain experiments and solid red lines represent model prediction using damping function parameters that are optimized to reproduce the Lissajous-Bowditch curves at 0.1 Hz. The maximum oscillatory stress amplitude is indicated on each subplot in units of Pa.

**Modeling of LAOS behavior:**

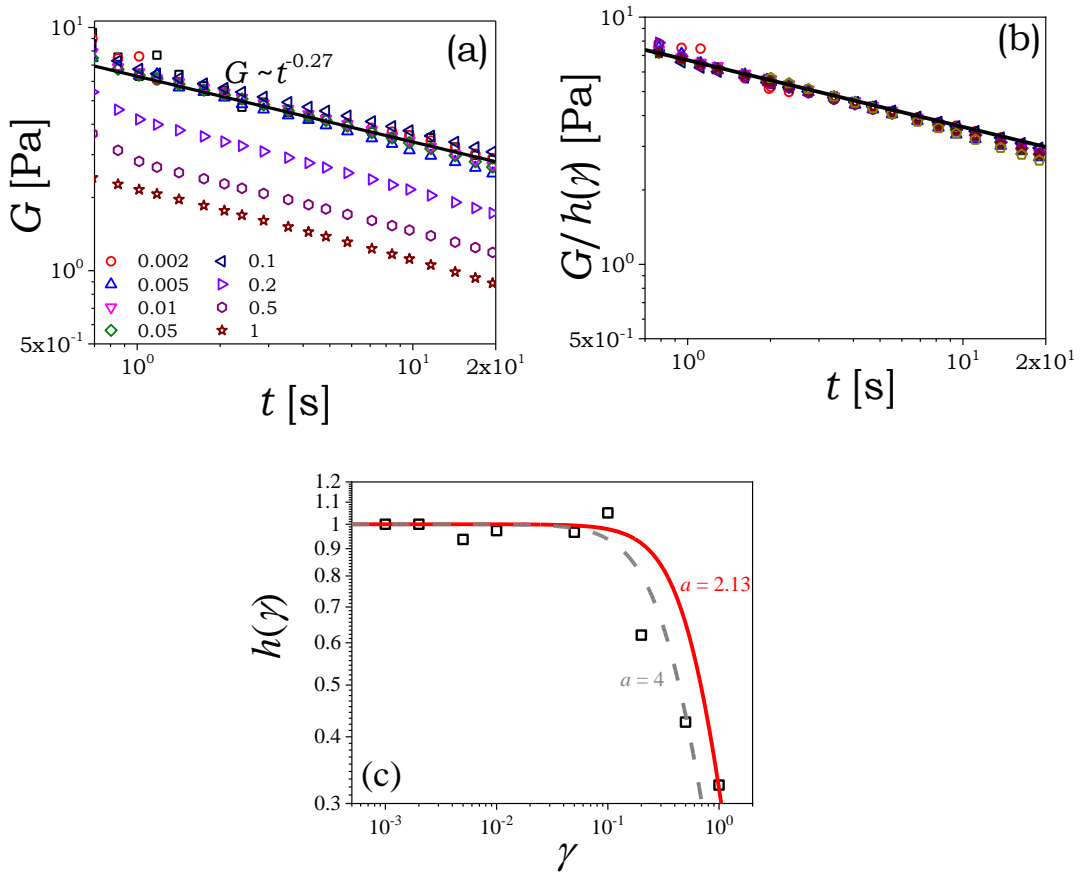

**Figure 8.** (a) Stress relaxation modulus ($G$) is plotted as a function of time after application of step strain of varying magnitudes mentioned in the legend at the critical gel state. The solid line denotes a power-law decay indicated by Eq. (12). (b) Vertically shifted stress relaxation modulus

is plotted as a function of time. (c) The inset shows the evolution of damping function $h(\gamma)$ as a function of strain amplitude. The dashed gray line denotes the fit of Eq. (14) with $a = 4$ and solid red line shows the fit with $a = 2.13$.

To model the behavior of the critical gel state, we conducted stress relaxation experiments over a strain range of $0.002 - 1$. A freshly prepared sample is used in each stress relaxation experiment, and the step strain is applied at the point of critical gel transition. The resulting decay of the stress relaxation modulus ($G$) is shown in Figure 8 (a). In the linear viscoelastic limit, $G$ for a critical gel exhibits a power-law dependence on time given by:[67]

$$G(t) = St^{-n_c}, \tag{12}$$

where $S$ is the gel strength. Consequently, the experimental relaxation modulus data is independent of the amplitude of strain within the linear viscoelastic region ($\gamma \leq 0.1$), and overlap on top of each other. The corresponding power-law index is identical to the value of $n_c$ obtained from the oscillatory studies as plotted in Figure 8 (a). Interestingly, as shown in Figure 8 (a), for the applied step strains beyond the linear regime, the stress relaxation curve obeys the same power-law dependence given by Eq. (12). However, the magnitude of the relaxation modulus decreases with an increase in $\gamma$. Beyond a certain threshold magnitude of step strain, power-law decay is not conserved, and we neglect that data in the current analysis. The strain-dependent stress relaxation modulus is represented as $G(t,\gamma)$. The self-similar behavior in $G(t,\gamma)$ over the explored duration of $\gamma$ allows vertical shifting using a damping function $h(\gamma)$. This leads to the superposition in $G(t,\gamma)$ as shown in Figure 8 (b). The existence of the superposition in $G(t,\gamma)$ suggests that the strain-dependent relaxation modulus can be factorized into time-dependent $G(t)$ and strain-dependent $h(\gamma)$ as:[52]

$$G(t,\gamma) = G(t)h(\gamma). \tag{13}$$

The vertical shift factor $h(\gamma)$ is plotted as a function of strain in Figure 8 (c). As expected, the value of $h(\gamma)$ is seen to remain close to 1 in the linear viscoelastic region. With an increase in the amplitude of strain, the value of $h(\gamma)$ decreases below 1. In the literature a number of analytical expressions of exponential, sigmoidal and other types have been proposed for $h(\gamma)$.[68] The data shown in Figure 8 (c) can be described by a popular functional form of damping function given by:[69-73]

$$h(\gamma) = \frac{1}{1 + a\gamma^2}, \tag{14}$$

where $a$ is the model parameter. This functional form captures the deviation of damping function from unity at increasing strain beyond linear viscoelastic limit and closely approximates the predicted behavior of damping function given by the Doi-Edwards reptation theory.[73, 74] Fitting Eq. (14) to the data in Figure 8 (c) yields $a = 4$ which describes the experimental data well as shown by the dashed line. This damping function is valid up to $\gamma \approx 1$, based on the range of experimental data in Figure 8 (b).

To model the nonlinearity, we employ the time-strain separable (TSS) K-BKZ (Kaye–Bernstein–Kearsley–Zapas) model. The integrand in the K-BKZ model can be factored into two components: the memory function $m(t - t') = \partial G(t - t')/\partial t'$, and a purely periodic part $h[\gamma(t) - \gamma(t')](\gamma(t) - \gamma(t'))$, where $\gamma(t) - \gamma(t') = \gamma_0(\sin \omega t - \sin \omega t')$ is the strain *accumulated* between times $t$ and $t'$. The shear stress is given by:[18]

$$\sigma(t) = \int_{-\infty}^{t} \frac{dG(t - t')}{dt'} h[\gamma(t) - \gamma(t')](\gamma(t) - \gamma(t')) dt'. \tag{15}$$

The nonlinear response stems from the damping function. In the limit of small values of accumulated strain, the damping function is close to unity, and the time-strain separable K-BKZ equation becomes equivalent to the Boltzmann superposition equation. We incorporate the damping function given by Eq. (14) into the K-BKZ equation for an oscillatory strain input $\gamma(t') = \gamma_0 \sin \omega t'$ to give the quasilinear constitutive equation:

$$\sigma(t) = S \int_{-\infty}^{t} \frac{d(t-t')^{-n_c}}{dt'} \left( \frac{1}{1+a\gamma^2} \right) (\gamma(t) - \gamma(t')) dt'. \tag{16}$$

Since all the material constants ($S$, $n_c$ and $a$) are known from LVE and step-strain experiments, the integral in Eq. (16) can be numerically solved for $\sigma(t)$ once the strain amplitude and frequency are specified. However, on application of oscillatory shear, numerical integration of TSS K-BKZ equation given by Eq. (16) becomes complicated due to the infinite domain of the integration and highly oscillatory nature of the strain at large frequencies.[75]

Recently, we developed a spectrally accurate method for solving the time-strain separable K-BKZ model in LAOS,[75] which has the same form given by Eq. (15). The efficiency and accuracy of the spectral method over other methods of numerical integration has already been established.[75] The spectral method takes care of the infinite limit of integration and oscillatory integrand without any uncontrolled assumptions. By changing variable $s = t - t'$, Eq. (15) can be rewritten as:[75]

$$\sigma(t) = \int_{-\infty}^{t} m(t-t') h[\gamma(t) - \gamma(t')] (\gamma(t) - \gamma(t')) dt' = \int_{0}^{\infty} m(s) p(\omega s; \omega t, \gamma_0) ds, \tag{17}$$

where $p(\omega s; \omega t, \gamma_0) = h[\gamma(t) - \gamma(t')](\gamma(t) - \gamma(t'))$ is a smooth periodic function of $\omega s$ with a period $2\pi/\omega$. We can accurately approximate $p(\omega s)$ with a discrete Fourier series (DFS), $S_{n_f}(\omega s)$, where the subscript $n_f$ represents the highest harmonic resolved in the Fourier series:[75]

$$p(\omega s) \approx S_{n_f}(\omega s) = \frac{a_0}{2} + \sum_{k=1}^{n_f} a_k \cos(k\omega s) + \sum_{k=1}^{n_f} b_k \sin(k\omega s). \tag{18}$$

In practice, $n_f$ can be increased and automatically selected to ensure that the maximum deviation $|p(\omega s) - S_{n_f}(\omega s)|$ is below a pre-specified error tolerance, such as $10^{-8}$ used here. The DFS coefficients can be efficiently determined using fast Fourier transform which has an asymptotic computational cost of $\mathcal{O}(n_f \log n_f)$. When $s = t - t' = 0$, $\gamma(t) - \gamma(t') = 0$, which implies $p(\omega s = 0) = h[\gamma(t) - \gamma(t')](\gamma(t) - \gamma(t')) = 0$, and hence $a_0/2 = -\sum_{k=1}^{n_f} a_k$ from Eq. (18). Using this

relation, and substituting Eq. (18) into Eq. (17), we obtain an approximation for the shear stress $\sigma(t) \approx \sigma_{n_f}(t)$ as:[75]

$$\sigma_{n_f}(t) = \sum_{k=1}^{n_f} \left( b_k \int_0^\infty m(s)\sin(k\omega s)ds - a_k \int_0^\infty m(s)(1-\cos(k\omega s))ds \right) \tag{19}$$

Interestingly, the dynamic moduli in the LVE regime are related to the memory function via:[76]

$$G'(\omega) = \int_0^\infty m(s)(1-\cos \omega s)ds = \frac{\pi S}{2\Gamma(n_c)} \omega^{n_c} \operatorname{cosec}\left(\frac{n_c \pi}{2}\right) \text{ and}$$

$$G''(\omega) = \int_0^\infty m(s)\sin \omega s\, ds = \frac{\pi S}{2\Gamma(n_c)} \omega^{n_c} \sec\left(\frac{n_c \pi}{2}\right), \tag{20}$$

where the second equality in both expressions above are the dynamic moduli for critical gels. This allows us to analytically evaluate the shear stress as:

$$\sigma_{n_f}(t) = \sum_{k=1}^{N} \left( b_k G''(k\omega) - a_k G'(k\omega) \right). \tag{21}$$

The only approximation invoked in this method is the representation of $p(\omega s)$ with a DFS given by Eq. (18). Consequently, the convergence of this method is spectral, and is insensitive to large values of $\omega$, which result in highly oscillatory integrals that degrade the performance of standard quadrature methods. Predictions using this spectral method, which avoids numerical quadrature, and conveniently side-steps the aforementioned singularity in the integrand, are shown by dashed gray lines in Figures 7. At the smallest strain amplitudes, the agreement between these predictions and experimental data is excellent. The model prediction of the stress response is perfectly elliptical in the linear domain and fits the experimental data well at all explored frequencies as seen in Figure 7. As strain amplitude increases beyond $\gamma_0 = 0.189$, predictions start to deviate systematically from the experimental data. Although the predicted stress response also exhibits broadening at the center, it underestimates its magnitude. Thus, we attribute most of the discrepancy between model predictions and experiments to the uncertainty in estimating the damping function, and the breakdown of the assumption of time-strain separability. As mentioned earlier, beyond a certain strain amplitude, the power-law behavior of the experimentally obtained $G(t,\gamma)$ breaks down. Therefore, like most real materials, the critical gel empirically exhibits TSS over a finite strain amplitude window. However, the TSS K-BKZ equation imagines an ideal material which obeys time-strain separability at all strain amplitudes.

Nevertheless, it is perhaps still meaningful to ask the question, "if we assume the material to be time-strain separable, what damping function is implied by the Pipkin diagram shown in Figure 7?" In other words, can we find parameters for the damping function that yield a more satisfactory fit of the experimentally observed Lissajous-Bowditch curves? Due to the efficiency of the spectral method, we can pose this as a least squares optimization problem by considering only a subset of the Lissajous-Bowditch curves in Figure 7. Here, we consider the data corresponding to the lowest frequency (0.1 Hz), i.e. the left-most column of the Pipkin diagram, and seek model parameter ($a$) that simultaneously provides the best fit to all the Lissajous-Bowditch curves in this subset. This exercise yields the optimized value for the parameter: $a = 2.13$ which is less than the value ($a = 4$) obtained from the step-strain experiment. The value of $a$ has been analyzed in the literature to relate to branching in polymers, and a smaller value is reported for branched topologies compared to linear polymers.[68,77] The modified damping function with smaller value of model parameter $a = 2.13$ is, therefore, suggestive of a denser percolated structure than reflected by the step strain measurements. A careful study combining rheology and scattering at a high strain field might be helpful in rendering insights about the microstructure.

The damping function with the modified parameter is shown as a solid red line in Figure 8 (c). Fits (0.1 Hz) and predictions (0.5 and 5 Hz) obtained using these optimized parameters and the spectral method are shown by solid red lines in Figure 7. Interestingly, the damping function inferred from a single frequency works remarkably well at other frequencies as indicated by solid red lines in Figure 7. It is worthwhile to mention that the prediction from the TSS K-BKZ equation with modified damping function matches the distortion in the experimental Lissajous-Bowditch curves exceedingly well in the nonlinear region. This is the first report on accurately modeling the nonlinear behavior of a critical gel by solving a simple TSS K-BKZ equation with modified damping function. Similar approach was previously reported for polymeric liquids by Giacomin *et al.*[14] They also found that TSS K-BKZ model using the damping function determined from step-strain experiments produced poor predictions of the LAOS response. However, fitting the damping function to a subset of the LAOS data produced a satisfactory description of the response at other frequencies and strain amplitudes.

**Microstructure under LAOS:**

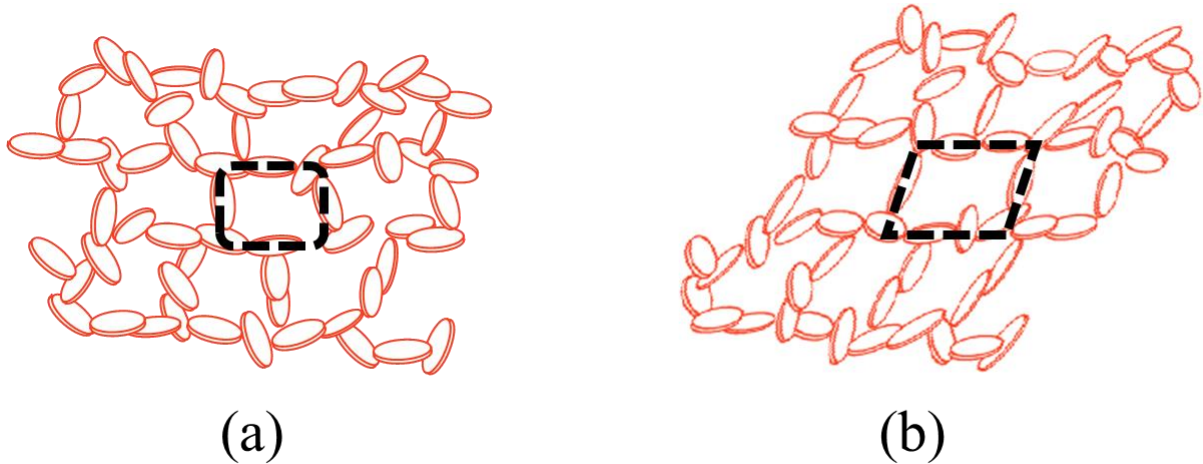

**Figure 9.** (a) Schematic representation of the percolated network structure in the critical gel of the colloidal dispersion. A single Hectorite particle is represented by the disk. (b) The microstructural changes occurring in the critical gel on the application of the strain amplitude in the nonlinear domain. The dashed rectangle highlights the structural changes discussed in the text.

The analysis of the large amplitude oscillatory shear and the resultant nonlinear parameters obtained from the transient data suggest that the studied material undergoes intracycle strain hardening, intracycle shear thinning and overall strain softening. In this section, we propose the possible microstructural changes occurring in the critical gel state leading to above-mentioned phenomena. The schematic represented in Figure 9 corresponds to a percolated network formed by the edge-face association between the colloidal particles. The percolated network remains unaltered at low amplitudes of strain and the response is viscoelastic without the presence of any nonlinear effects. The dashed rectangle highlights a particular section of the network, which we shall analyze when subjected to strain in the nonlinear region. The estimated elastic nonlinear parameters ($G'_M$ and $G'_L$) suggest intracycle strain hardening, while the viscous parameters ($\eta'_M$ and $\eta'_L$) indicate intracycle shear thinning in the system. We believe that the intracycle strain hardening is likely due to deformation of the network of clay particles, which offers greater resistance to the elastic deformation due to the stretching of the same. The dashed rectangle in Figure 9 representing the interparticle associations can be seen to get stretched upon the application of nonlinear strain deformation. Furthermore, the particles also get aligned in the direction of flow as indicated by the horizontal shift of the highlighted rectangle in Figure 9 (b), which results in intracycle shear thinning. At high amplitudes in the nonlinear region, the network junctions might break, and the broken clay clusters have little chance to rejoin the network structure under continuous deformation. This results in overall strain thinning behavior in the critical gel as evident by the decrease of moduli in a regular strain sweep test. Although it may seem paradoxical to have intracycle strain hardening and overall shear thinning simultaneously in a system, it is important to note that LAOS analysis of a single oscillation cycle suggesting strain hardening is a local effect while the regular strain sweep suggesting strain thinning is an overall behavior.[78]

This work, therefore, presents a comprehensive characterization of a critical gel in the linear as well as in nonlinear domain. The TSS K-BKZ model describes the rheological behavior of the critical gel ranging from the linear to nonlinear domain. The damping function drives the nonlinear response of the model. Remarkably, the TSS K-BKZ model with optimized damping function parameters solved using an efficient numerical method describes the experimental data well. This work is the first report on the experimental characterization and theoretical modeling of LAOS behavior of a critical gel. The proposed model is the first to describe the complete rheological behavior of the critical gel from linear to nonlinear domain. We believe that the simple model proposed in this work can be extended to a wide range of soft materials at the critical gel state to study the linear, weakly nonlinear, and nonlinear viscoelastic behavior.

## 5. Conclusions:

We investigate the nonlinear behavior of synthetic hectorite clay at the critical gel state by estimating higher-order elastic and viscous moduli. Inspection of the intracycle data using Lissajous-Bowditch curves renders physical insights about the nonlinearity. The dominance of large-strain modulus over minimum-strain modulus suggests intracycle strain hardening in the critical gel. On the other hand, a higher value of viscosity at minimum strain rate than at highest strain rate indicates intracycle shear thinning in the system. We believe that the extension of the network of clay particles offers more resistance to elastic deformation which manifests in intracycle strain hardening. Furthermore, as the network gets strained, the particles get aligned in the direction of shear, which results in intracycle shear thinning. In order to model the linear-nonlinear transition, we perform a stress relaxation experiment on the critical gel state at different magnitudes of strains. Up to a certain magnitude of strain, the vertical shifting of the independent stress relaxation curves leads to a superposed master curve. The corresponding strain-dependent vertical shift factor, also known as a damping function, is observed to decay in a quadratic fashion with strain beyond the linear region. On incorporating the damping function into the time strain separable K-BKZ integral equation leads to a quasilinear constitutive equation. We solve the TSS K-BKZ model on the application of an oscillatory strain using the spectral method. Although the proposed model predicts the intracycle stress response quite well in the linear as well as the nonlinear domain, a slight modification in the damping function leads to enhanced accuracy in the prediction of the experimental data. This leads to a simple formulation to analyze the nonlinear response of a critical gel. Furthermore, the qualitative features, including the intracycle strain hardening and intracycle shear thinning, remain preserved by the developed viscoelastic model.

**Supplementary Material:**

See the supplementary material for the analysis of nonlinear behavior in terms of Chebyshev coefficients.

**Acknowledgements:**

K.S and Y.M.J acknowledge the financial support provided by Science and Engineering Research Board (SERB), Department of Science and Technology, Government of India. S.S acknowledges the support by National Science Foundation under Grant No. NSF DMR-1727870. The authors are grateful to Professor Randy H. Ewoldt for providing us with the MITlaos software package.

**Data availability statement:**

The data that support the findings of this study are available from the corresponding author upon reasonable request.

# Supplementary Material

## Large Amplitude Oscillatory Shear Study of a Colloidal Gel at the Critical State

Khushboo Suman,[1a] Sachin Shanbhag,[2*] and Yogesh M. Joshi[1*]

[1]Department of Chemical Engineering, Indian Institute of Technology Kanpur, Kanpur 208016, India

[2]Department of Scientific Computing, Florida State University, Tallahassee, Florida 32306

[a]Present address: Department of Chemical & Biomolecular Engineering, University of Delaware, Newark, DE 19716 USA

* Author to whom correspondence should be addressed; electronic mail: sshanbhag@fsu.edu and joshi@iitk.ac.in

**Analysis of nonlinear behavior in terms of Chebyshev coefficients**

The unknown material properties in the stress decomposition equation given by Eq. (5) and (6) of the main manuscript can be replaced by Chebyshev polynomials of the first kind. The advantages of Chebyshev polynomials over Laguerre and Hermite polynomials or Jacobi polynomials are symmetry about $x=0$ and orthogonality over a finite domain [-1,1].[1] Using Chebyshev polynomials of the first kind, the elastic and viscous contributions can be expressed as:

$$\sigma' = \gamma_0 \sum_{n,odd} e_n(\omega,\gamma_0) T_n(x), \text{ and} \tag{S1}$$

$$\sigma'' = \dot{\gamma}_0 \sum_{n,odd} v_n(\omega,\gamma_0) T_n(y), \tag{S2}$$

where $x = \gamma/\gamma_0$, $y = \dot{\gamma}/\dot{\gamma}_0$, $T_n(x)$ is the $n^{th}$ order Chebyshev polynomial, $e_n$ is the $n^{th}$ order elastic Chebyshev coefficient with units of modulus and $v_n$ is the $n^{th}$ order viscous Chebyshev coefficient with units of viscosity. The Chebyshev coefficients are related to the Fourier coefficients by:[1]

$$e_n = G'_n (-1)^{(n-1)/2}, \qquad n: odd \tag{S3}$$

$$v_n = \frac{G''_n}{\omega} = \eta'_n, \qquad n: odd. \tag{S4}$$

In the limit of linear viscoelasticity, we get $e_3/e_1 \ll 1$ and $v_3/v_1 \ll 1$. Furthermore, the above equations clearly point out that the Chebyshev coefficients reduces to linear viscoelastic results: $e_1 = G'_1$ and $v_1 = \frac{G''_1}{\omega}$. With an increase in the strain amplitude, we approach the nonlinear regime, where the higher harmonics become significant. For $n=3$, a positive contribution from third-order elastic Chebyshev polynomial $T_3(x)$ suggests higher elastic stress than the first-order contribution at the maximum value of strain. Therefore, a positive value of $e_3$ signifies intracycle strain stiffening and negative value suggest intracycle strain softening. Similarly, $v_3 > 0$ corresponds to intracycle shear thickening and $v_3 < 0$ indicates intracycle shear thinning. The Chebyshev coefficients can also be related to the nonlinear moduli and viscosities by the following relations:[1]

$$G'_M = e_1 - 3e_3 + ... \tag{S5}$$

$$G'_L = e_1 + e_3 + \ldots . \tag{S6}$$

$$\eta'_M = v_1 - 3v_3 + \ldots , \tag{S7}$$

$$\eta'_L = v_1 + v_3 + \ldots , \tag{S8}$$

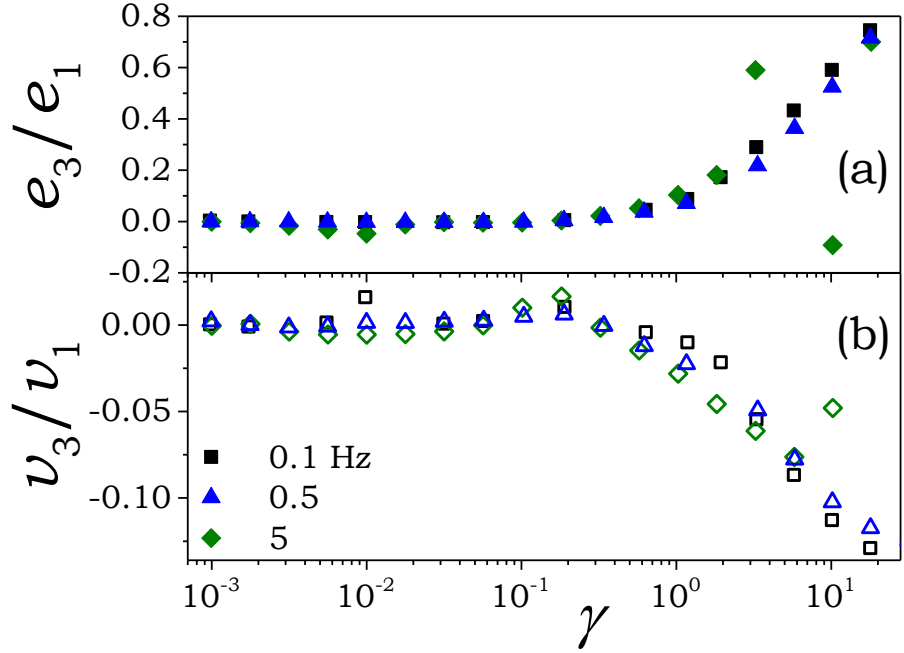

**Figure S1.** The variation of third-order (a) elastic Chebyshev coefficient ratio $(e_3/e_1)$, (b) viscous Chebyshev coefficient ratio $(v_3/v_1)$ is plotted as a function of strain amplitude at different frequencies.

To characterize the nonlinear behavior of critical gel, we plot the third-order elastic $(e_3/e_1)$ and viscous $(v_3/v_1)$ Chebyshev coefficient ratio as a function of strain amplitude at different values of explored frequencies in Figure S1 (a) and (b). Since the third-order Chebyshev coefficients are significantly larger than the higher-order coefficients, we focus on only the first and third order harmonics in this work. It can be seen that $e_3/e_1$ and $v_3/v_1$ values are close to zero in the linear viscoelastic region. The non-zero values of the third-order Chebyshev coefficient ratio is a manifestation of the nonlinear viscoelasticity. As strain amplitude increases above the critical value of 0.1, $e_3/e_1$ becomes positive, which corresponds to the intracycle strain stiffening behavior. Conversely, the value of $v_3/v_1$ becomes negative at higher magnitudes of strain indicating intracycle shear thinning behavior. The nonlinearity illustrated by the Chebyshev coefficient gets more enhanced with increase in the strain amplitude.[1] The qualitative behavior of $e_3/e_1$ and $v_3/v_1$ remains preserved at all the values of explored frequencies. Therefore, the observed behavior suggests that the critical gel state of studied clay dispersion undergoes intracycle strain stiffening and intracycle shear thinning behavior at all the explored frequencies. The identical intracycle behavior is also inferred from nonlinear moduli and viscosities as discussed in the main manuscript file.